\documentclass[letter,structabstract]{aa}  
\usepackage{graphicx}
\usepackage{txfonts}
\begin{document}
   \title{High resolution spectroscopy of Pluto's atmosphere: detection of the 2.3 $\mu$m CH$_4$ bands
and evidence for carbon monoxide}
%

   \author{E. Lellouch
          \inst{1}
	  \and
          C. de Bergh\inst{1}
          \and
          B. Sicardy\inst{1,2}
	  \and
	  H.U. K\"aufl\inst{3}
	  \and
	  A. Smette\inst{4}
         }

   \institute{LESIA, Observatoire de Paris, 5 place Jules Janssen, 92195 Meudon, France;
              \email{emmanuel.lellouch@obspm.fr}
\and 
Universit\'e Pierre et Marie Curie, 4 place Jussieu, F-75005 Paris, France;  senior member of the Institut Universitaire de
France
         \and	
            European Space Observatory, Karl-Schwarzschild-Strasse 2, D-85748 Garching bei M\"unchen, Germany  
         \and	
            European Space Observatory, Alonso de C\'ordova 3107, Casilla 19001, Vitacura, Chile 
             }

   \date{Received March, 24, 2011; revised, April 18, 2011; accepted, April 21, 2011}

 
  \abstract
{}
   {The goal is to determine the composition of Pluto's atmosphere and to constrain the nature of surface-atmosphere interactions.}
   {We perform high--resolution spectroscopic observations in the 2.33--2.36 $\mu$m range, using CRIRES at the VLT.}
   {We obtain (i) the first detection of gaseous methane in this spectral range, through lines of the $\nu_3$ + $\nu_4$
and $\nu_1$ + $\nu_4$ bands (ii) strong evidence (6-$\sigma$ confidence) for gaseous CO in Pluto. For an isothermal atmosphere at 90 K, the CH$_4$
and CO column densities are 0.75 and 0.07 cm-am, within factors of 2 and 3, respectively. Using a physically--based thermal structure
model of Pluto's atmosphere also satisfying constraints from stellar occultations, we infer CH$_4$ and CO mixing ratios q$_{CH_4}$= 0.6$^{+0.6}_{-0.3}$ \% (consistent with results from the 1.66 $\mu$m
range) and q$_{CO}$ = 0.5$^{+1}_{-0.25}$$\times$10$^{-3}$. The CO atmospheric abundance is consistent with its surface abundance.
As for Triton, it is probably controlled by a thin, CO-rich, detailed balancing layer resulting from seasonal transport and/or atmospheric escape. }
   {}

   \keywords{Solar system:general ; Infrared: solar system ; Pluto}  
 
\titlerunning{Evidence for CO in Pluto's atmosphere} 
   \maketitle
%

\section{Introduction}
Along with Triton's, Pluto's tenuous atmosphere is a benchmark example of an atmosphere controlled by vapor pressure equilibrium
with surface ices. Pluto's surface is dominated by N$_2$ ice, but other detected surface compounds include CH$_4$, CO and,
tentatively, C$_2$H$_6$ and other non--volatile species (e.g. Cruikshank et al. 1997, Dout\'e et al. 1999, Grundy
and Buie 2001, 2002, Olkin et al. 2007, Protopapa et al. 2008, Merlin et al. 2010). Pluto's atmospheric composition remains
poorly known. Based on its surface abundance and large volatility, and albeit not observed spectroscopically, 
N$_2$ is known to dominate Pluto's atmosphere. The only other gas phase species to have been detected is methane.
It was first observed in its 2$\nu_3$ band near 1.66 $\mu$m  by Young et al. (1997), providing the first estimate of its 
CH$_4$ column density (1.2 cm-am within a factor of 3-4, assuming a gas temperature of 100 K). 
Much improved observations obtained with the CRIRES infrared echelle spectrograph at the VLT in 2008 (Lellouch et al. 2009), including
the detection of numerous lines, up to J~=~8, made it possible to separate temperature and abundance effects in the Pluto spectra,
and to constrain the depth of Pluto's atmosphere. Assuming an isothermal atmosphere, the data indicated a mean gas temperature 
T = 90$^{+25}_{-18}$ K and a methane column density a$_{CH_4}$ = 0.75$^{+0.55}_{-0.30}$ cm-am. 
Although Pluto's surface pressure is uncertain, the
 combined analysis of these data with stellar occultation curves indicated a CH$_4$/N$_2$
mixing ratio q$_{CH_4}$=0.5$\pm$0.1~\%. 


The other species to be expected in Pluto's atmosphere based on volatility considerations is carbon monoxide. 
Searches for CO have been conducted already a decade ago 
(Bockel\'ee-Morvan et al., 2001, Young 
et al. 2001), but as detailed below, results have been relatively unconstraining.  More recently (July 2009), and using again CRIRES/VLT,
Lellouch et al. (2010) detected CO in Triton's atmosphere from its CO(2-0) band at 2.35~$\mu$m, with a column density of $\sim$0.30 cm-am,
corresponding to CO/N$_2$ $\sim$ 6$\times$10$^{-4}$ for an estimated 40 $\mu$bar surface pressure. They also searched for CO on Pluto 
in the (3-0) band near 1.57 $\mu$m. The choice of that band was motivated by the much larger brightness of Pluto in H vs K band 
(geometric albedo $\sim$0.6 at 1.57 $\mu$m vs $\sim$0.2 at 2.35~$\mu$m).
On the other hand, lines of the CO(3-0) band, whose strengths are  $\sim$2 orders of magnitude weaker than those of the (2-0) band, are expected
to show up only for large CO abundances, and as a matter of fact, only an upper limit of 1 cm-am was obtained, i.e.  
CO/N$_2$ $<$ 5$\times$10$^{-3}$ for a typical 15 $\mu$bar pressure. Based on considerations on the 
surface/atmosphere equilibrium and the CO surface abundance, estimates of the CO abundance on Pluto are $\sim$ 10 times smaller. 
Reaching this sensitivity in the near-IR demands the use of CO(2-0) band, but given that Pluto is $\sim$3.5 times 
fainter than Triton at 2.35 $\mu$m, this in turn requires a deep integration. We here report on such observations.

    \begin{figure*}
   \centering
   \includegraphics[width=15.4cm,angle=0]{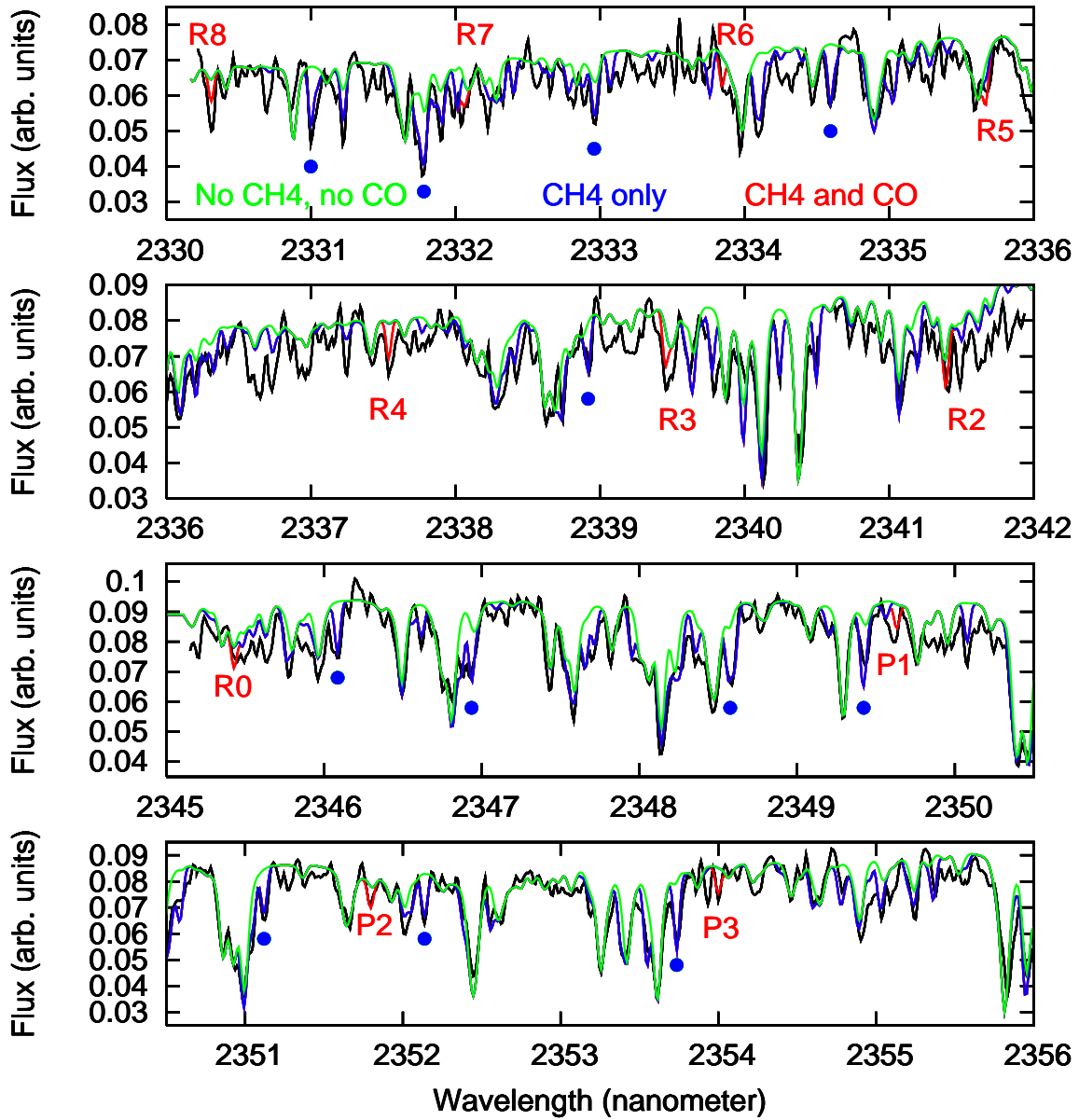}
   \caption{Black: Pluto spectrum at  2330-2342 and 2345-2356 nm (arbitrary units). The total integration time is 7h20 min. The spectral resolution is 60,000. Red, blue and green curves show synthetic spectra for an isothermal Pluto atmosphere at T = 90 K,
and including telluric and solar lines. Green: no methane and CO in Pluto's atmosphere. Blue: methane is included with a 0.75 cm-am
column density. Red: CO is also included with a column density of 0.07 cm-am. Blue dots locate the methane features that were used in the line coaddition process shown in Fig. 2. }
              \label{FigCH$_4$}%
    \end{figure*}

\section{VLT/CRIRES observations}
Spectroscopic observations of Pluto were obtained during two half--nights on July 27 (0h40-4h25 UT) and 29 (0h10-5h16 UT) 2010, using the CRIRES 
instrument (K\"aufl et al. 2004) installed on ESO VLT (European Southern Observatory Very Large Telescope) UT1 (Antu) 8.2 m telescope.
We used the instrument in adaptive optics mode and with a slit of 0.4". The spectral resolution, measured from the linewidths and the source effective size, was 60,000. We targeted a portion
of the (2-0) band of CO, also encompassing a fraction of the $\nu_1$ + $\nu_4$ and $\nu_3$ + $\nu_4$ bands of CH$_4$. The instrument includes four detectors. On July 27, we covered the 2312-2325, 2330-2342, 2345-2356 and 2359-2370 nm ranges. However, detectors 1 and 4 were affected by flux losses along the slit. On July 29, we used a fast (``windowed") readout  mode in which only detectors 2 and 3 are read (and windowed), resulting in a noticeable gain in sensitivity for faint targets. Pluto's mean (East) longitude for the two dates was 118$^{\circ}$ and 5$^{\circ}$, respectively. A large Doppler shift ($\sim$15.9 and $\sim$16.7 km/s for July 27 and 29) ensured proper separation of the target Pluto lines from the telluric absorptions, particularly from CH$_4$. Each night we observed a telluric standard (HR 5917) to check for wavelength calibration and atmospheric transmission. All data were reduced using the standard steps of the CRIRES pipeline, including corrections for darks, flatfield, 
image recombination, replacement of bad pixels and outliers, and spectral extraction.
Spectra from the two dates were finally coadded, providing a mean S/N of 15-20 per resolving element.




   \begin{figure*}
   \centering
   \includegraphics[angle=-90,width=15.5cm]{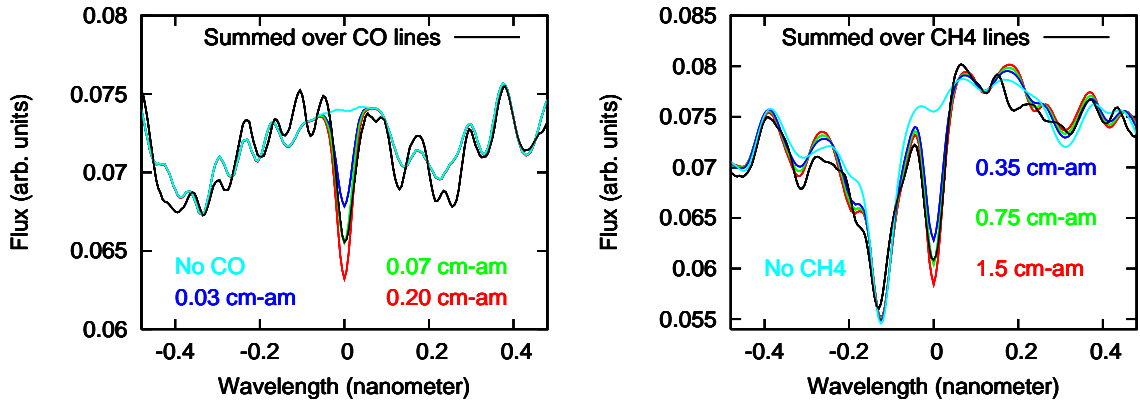}
   \caption{Black: Coadded CO (left) and CH$_4$ (right) line, obtained from coaddition of the observed spectrum within $\pm$0.5 nm
of (i) the R2-R7 and P1-P3 lines of CO and (ii) the 12 CH$_4$ features identified by dots in Fig. 1. Colored lines: result of the same treatment applied to synthetic spectra. 
In the right panel, the feature at --0.13 nm is mostly due to the telluric counterpart of the coadded methane line.}
              \label{FigCH$_4$}%
    \end{figure*}

\section{Modelling and evidence for CO}
The total spectrum in the 2330-2342 and 2345-2356 nm ranges (Fig. 1) shows the detection of many features due to the $\nu_3$ + $\nu_4$ band of CH$_4$
(plus some lines of the $\nu_1$ + $\nu_4$ band longwards of 2350 nm) representing the first detection of these bands in Pluto's atmosphere.  As in our previous studies (Lellouch et al. 2009, 2010)  we constructed a direct, disk-integrated, line-by-line atmospheric model of Pluto, including solar lines  
reflected off Pluto's surface as well as the telluric transmission and accounting for the proper Doppler  shifts. In a first step, we assumed a single, homogeneous layer at 90$^{+25}_{-18}$ K,
as inferred from the 1.66 $\mu$m spectrum (Lellouch et al. 2009). This yielded a best fit methane column density a$_{CH_4}$ = 0.75$^{+1.0}_{-0.5}$  cm-am,
consistent with the 2$\nu_3$ inferences. Here the large error bar combines quadratically a factor-of-2 uncertainty  due to the S/N limitations and a similar uncertainty  due to the range in mean gas temperature. 
The lower precision compared to results from the 1.66 $\mu$m band is due to the lower S/N of the 2.35 $\mu$m spectrum. 

Pluto's thermal structure is characterized by a ``warm" ($\sim$100~K) upper atmosphere, a sharp inversion layer in the 3-10~$\mu$bar pressure
range, 
and possibly but not necessarily a cold ($\sim$ 36 - 45 K) troposphere. By combining their 1.66~$\mu$m CH$_4$ spectrum with 
occultation data, Lellouch et al. (2009) 
found that the stratospheric gradient is in the range 3--15 K/km, the (current) surface pressure is 6.5--24 $\mu$bar, and the troposphere is at most 17 km deep. Each of the allowed thermal profiles is associated with a methane column density a$_{CH_4}$, required to fit
the 1.66~$\mu$m spectrum. Deeper (i.e. colder) models require larger methane columns than shallower models. The range for a$_{CH_4}$
in these models was then 0.65--1.3 cm-am, but 
the CH$_4$ mixing ratio could be accurately determined to q$_{CH_4}$=0.5$\pm$0.1 \%.
We re--modelled the 2.35 $\mu$m spectrum with one ``nominal"
thermal profile, having a stratospheric gradient of 6 K/km, a wet tropospheric adiabat, and a surface pressure of 15 $\mu$bar,
corresponding to a N$_2$ column density of 210 cm-am (this profile is one of the ``green" profiles shown in Fig. 4 of Lellouch
et al. 2009). Assuming that methane is vertically uniform, an identical fit to that shown in Fig. 1 is then obtained for a$_{CH_4}$ = 1.2 cm-am, i.e. 
q$_{CH_4}$ = 0.57 \%.  Refitting the data with atmospheres restricted to $p$~=~6.5~$\mu$bar or extending to $p$~=~24 $\mu$bar (Table 1) indicates
that S/N limitations dwarf thermal profile uncertainties, 
and the methane mixing ratio indicated by our
spectrum is 0.6$^{+0.6}_{-0.3}$ \%.

Besides the CH$_4$ features, Pluto's atmosphere spectrum shows absorptions at the positions of most of the CO(2-0) lines covered by the observed
spectral range (R8---R2, R0 and P1-P3), and that are not fit by the previous methane-only model. Including CO at 90 K and with a column density
of 0.07 cm-am clearly improves the fit in the vicinity of all these lines. Admittedly, several mismatches remain, that are comparable to or
sometimes even larger than the features tentatively identified as due to CO; examples occur near 2330.6, 2336.6, 2337 or 2353.9 nm. To assess
the significance of the result, we coadded the data over $\pm$0.5 nm broad intervals centered on all the CO line positions, without
any a priori exclusion (except for the R0 and R8 lines that are too close to detector edge).  This treatment (Fig. 2, left) produces 
clear evidence for an absorption line at the ``zero wavelength", around which significant structure is observed. Performing
the same treatment on synthetic spectra allows us to interpret this residual structure as being due to the more or less random co-addition of other lines (telluric, solar, or Pluto CH$_4$) in the vicinity of the CO lines. The remaining standard deviation between model and observation 
($\sim$1.3$\times$10$^{-3}$ in the units of Fig. 2, i.e. 1.8 \% of the continuum) indicates a 6-$\sigma$ detection of the ``combined
CO line". For an isothermal 90~K atmosphere, the best fit is achieved for a CO column density a$_{CO}$  = 0.07 cm-am. For the above nominal thermal
profile, it is obtained for a$_{CO}$~=~0.10~cm-am, i.e. q$_{CO}$ = 4.8$\times$10$^{-4}$, assuming uniform mixing. 
Given the sensitivity of the models to the CO abundance -- and the small impact of thermal profile uncertainties (Table 1) -- the allowed range is 
q$_{CO}$ = 5$^{+10}_{-2.5}$$\times$10$^{-4}$. Although the S/N of the Pluto spectrum is considerably lower than
that for Triton (Lellouch et al., 2010), the accuracy on the CO abundance is comparable. This stems from the fact that the Pluto CO
lines are instrinsically less narrow and saturated than their Triton counterparts, given the higher gas temperature ($\sim$90 K vs $\sim$50 K) and $\sim$4 times smaller CO column density on Pluto vs Triton. 

\begin{table}
\caption{Best fit CH$_4$ and CO abundances vs surface pressure.}\label{starbursts}
\centering
\begin{tabular}{lcccc}
\hline 
Pressure      &     \multicolumn{2}{c}{CH$_4$} &     \multicolumn{2}{c}  {CO}\\
($\mu$bar)         &    Col. dens.&  Mixing ratio  &    Col. dens.   & Mixing ratio \\
6.5        &             0.6 cm-am      &      6.0$\times$10$^{-3}$  &           0.07 cm-am        &  7.0$\times$10$^{-4}$ \\
15         &              1.2 cm-am   &         5.7$\times$10$^{-3}$ &            0.10 cm-am    &     4.8$\times$10$^{-4}$\\
24           &             1.7 cm-am  &          5.3$\times$10$^{-3}$  &           0.14 cm-am   &      4.4$\times$10$^{-4}$\\
\hline
\end{tabular}
\end{table}


A similar line coaddition process can be performed on the CH$_4$ lines. However, the larger complexity of the methane spectrum implies some
a priori selection of the CH$_4$ lines. An example is shown in the right panel of Fig. 2, using the twelve CH$_4$ features marked in Fig. 1.
The ``combined CH$_4$ line" is then detected at $\sim$11-$\sigma$. With CH$_4$/N$_2$ = 0.6 \%,
the  CH$_4$/CO ratio in Pluto's atmosphere is nominally equal to 12, but could be in the range 2--48 for the CH$_4$ determined in this study,
or 2.5--24 if the more precise value q$_{CH_4}$ = 0.5$\pm$0.1 \% is used.

\section{Discussion}
The CO amount we infer is much smaller than values previously reported. Using the same CO(2-0) band and IRTF/CSHELL, Young et al. (2001) 
obtained a thermal-profile dependent upper limit a$_{CO}$ $<$ (1.2--3.5)$\times$10$^{21}$ cm$^{-2}$, i.e. $<$45-130 cm-am. This limit 
was improved by $\sim$2 orders of magnitude from our VLT/CRIRES observations of the CO(3-0) band (a$_{CO}$ $<$ 1 cm-am, Lellouch et al. 2010), 
but the present CO determination is another factor-of-10 lower. At millimeter wavelengths. Bockel\'ee-Morvan et al. (2001) obtained a tentative (formally 4.5~$\sigma$) detection of the CO(2-1) line at 1.3 mm, using the IRAM 30-m telescope, that, if real, would indicate a CO/N$_2$ mixing ratio in the 1.2--7 \% range (also thermal-profile dependent). However, it was more cautiously interpreted as an upper limit, due to strong galactic contamination in their Pluto spectrum.  The associated CO column was $<$3-7 cm-am,
over a factor of 6 more constraining than the Young et al. (2001) results. The present evidence for CO suggests that the authors were indeed
wise to regard their result as an upper limit.  

Pluto's surface pressure reflects sublimation equilibrium for N$_2$ ice, and N$_2$ may follow saturation in a putative Pluto troposphere.
Yet, the large CH$_4$ mixing ratio at $\sim$90 K in Pluto's stratosphere implies that it is not severely affected by atmospheric condensation, 
and that if Pluto has a troposphere, strong supersaturation of CH$_4$ occurs (Lellouch et al., 2009). Rannou and Durry (2009)
reached the same conclusion from microphysical arguments, but also mentioned that N$_2$ and CO can easily condense. Given its
equilibrium vapor pressure, only a factor of $\sim$8 lower than N$_2$ at 37 K, a $\sim$0.05 \% CO/N$_2$ mixing ratio
implies instead that CO condensation does not occur. 

As first pointed out by Lellouch (1994), CO is a cooling agent in Pluto's atmosphere through radiation in its pure rotational lines, but its importance has been initially overstated. Lellouch found that when CO cooling was included in the Yelle and Lunine thermal model (1989) -- that included absorption in the 3.3 $\mu$m band of CH$_4$ as the only energy source -- the $\sim$100 K stratospheric temperature required a very large ($>$~10~\%) CH$_4$ abundance. 
Strobel et al. (1996) showed that this requirement was relaxed to $\sim$ 3 \% when heating in the 1.6 and 2.3 $\mu$m  CH$_4$ bands was considered. Still,
in the presence of CO, a slightly negative temperature gradient (mesosphere) is expected in the sub-microbar region, 
evidence for which was claimed in occultation-derived thermal profiles (Elliot et al. 2007). However, 
the recent analysis by Zalucha et al. (2011) did not confirm the need for CO to explain occultation data. These 
authors directly modelled occultation light-curves with physically-based thermal profiles calculated from a radiative--conductive model inherited from Strobel et al. (1996), and in which the surface pressure and the CH$_4$ and CO mixing ratios are free parameters. Zalucha et al. (2011) deduced CH$_4$ mixing ratios of 0.18--0.94 \% for occultations recorded in 1988, 2002, 2006, and 2008 (albeit with no specific trend with time), broadly 
consistent with the observed $\sim$0.5~\% mixing ratio. In contrast, and although they confirmed that CO does affect the thermal structure 
in the $\mu$bar region (see e.g. their Fig. 8), their simulations indicated that occultation lightcurves are essentially insensitive to the CO mixing ratio. As a further testimony of the less than originally thought importance of CO for controlling Pluto's atmosphere state, Strobel (2008) found that CO plays at most a minor role in the atmospheric escape rates. 

Our inferred atmospheric CO mixing ratio is consistent with the long-standing predictions by Owen et al. (1993), Lellouch (1994) and Strobel et al. (1996). These estimates, which spanned the range (2--20)$\times$10$^{-4}$, were obtained assuming an ideal N$_2$--CO--CH$_4$ solid solution (Raoult's law) and using the then 
available CO/N$_2$ ice mixing ratio (0.5 \%, from Owen et al.). However a refined analysis of the near-IR spectra indicates that CO is present 
on Pluto's surface with a mixing ratio of 0.08--0.2 \% relative to N$_2$ (Dout\'e et al. 1999). Using modern vapor pressure
data (Fray et al. 2010), the ideal mixture case leads to an atmospheric CO/N$_2$ of (1.0--2.7)$\times$10$^{-4}$, 
only marginally in agreement with the observed (2.5--15)$\times$10$^{-4}$ range. Rather, the atmospheric
and surface CO/N$_2$ mixing ratios are the same within error bars, a situation predicted by the 
``detailed balanced" model (Trafton, 1990; Trafton et al. 1998). In this scenario, surface-atmosphere exchanges in presence of escape and seasonal transport lead to an atmospheric composition reflecting that of the accessible ice reservoir from which it is replenished. In the case,
relevant for a CO / N$_2$ mixture, where no fractionation occurs during escape or transport, the process equalizes the mixing ratios in the
atmosphere and the volatile reservoir. This is made possible by the formation of a thin CO-enriched surface veneer in
equilibrium with the atmosphere according to Raoult's law. Lellouch et al. (2010) found this scenario to be valid for CO on Triton.

The alternative scenario would be that the enhanced CO abundance compared to the ideal solution case results from the presence
of pure CO patches. The CO partial pressure is 4--25 nbar, i.e. 80--500 times less than the vapor pressure of pure CO. Hence,
patches of pure CO covering 0.2--1.2 \% of Pluto's surface could in theory produce the elevated CO abundance. Problems
with this scenario are that (i) the formation of pure CO grains is not expected thermodynamically, given the complete miscibility of N$_2$ and CO (ii) CO is not buoyant in N$_2$, inhibiting the sublimation of any CO patch (Stansberry et al. 1996). 
Yet, the longitudinal distribution of the CO 1.58 $\mu$m ice features shows evidence for a CO-rich
region near longitude L~=~180$^{\circ}$ (Grundy and Buie 2001), which (coincidentally or not) corresponds to the brightest and least red region in HST maps
near Pluto's equator.
(Buie et al. 2010). This situation contrasts with the Triton case, where the longitudinal distributions of the N$_2$ and CO ice bands
are remarkably similar  (Grundy et al. 2010). 
How could CO become concentrated in a localized spot on Pluto remains to be elucidated. Hopefully,
combined imaging and spectroscopy from New Horizons will reveal the nature of this CO--rich region and whether it may contribute
to the CO atmospheric abundance.

\begin{acknowledgements}
This work is based on observations performed at the European Southern Observatory (ESO), proposal 085.C-0113. 
\end{acknowledgements}

\end{document}